\documentclass[letterpaper, 10 pt, conference]{ieeeconf}  

\IEEEoverridecommandlockouts                              
\usepackage{graphicx} 
\usepackage{amsmath} 
\usepackage[]{algorithm2e}
\usepackage[bookmarks=false]{hyperref}
\usepackage{biblatex}
\title{\LARGE \bf
Optimal Group Formulation Using Machine Learning
}

\author{Mahbub Hasan$^{1}$ and Md. Al-Emran$^{2}$}

\begin{document}

\maketitle
\thispagestyle{empty}
\pagestyle{empty}

\begin{abstract}
Group formation itself a perplexing process. Over the decade of time education and others disciple has improved imminently but optimal group formation in educational system is still struggling. Our research focus on to create optimal group in a class of any institute. In this research we use Simulated Annealing (SA) for best group formation based on the previous academic record. We generally create an arbitrary cluster first then optimise using SA. Our model has significant success rate over a large number of datasets. This research will play a pioneer role in group formation`s in the academic and related researches.\\

\end{abstract}

Keywords: Optimal Group Formulation (OGF), Simulated Annealing (SA), Machine Learning (ML),

\section{INTRODUCTION}

At the present time, enormous quantity of data is available for producing essential information.  In medical, education, banking, business etc. institute can get beneficiary through the acquired information. For any civilization education treat as a backbone. The role for an educational institute is to teach and train every student so that they can accomplish their desire goal in the future. Despite having unvarying stage to learn in the schools/colleges, the performance of students differs hugely. This can be qualified to alteration in cognition level, motivation levels and environmental influences.\parencite[]{mahbub}

Optimal Group Formulation (OGF) has become an evolving zone for research interest among scientists and researchers. OGF translate which groups turns the most significant value in every aspect.
In this paper we proposed a machine learning based group formation model, which will generate a optimal solution based on each students previous record.

The main objective of this paper is to find out groups for students in a particular theory or lab class through our proposed model. The most significant contribution by this research, it will generate a group for future work and show the error rate.

The rest of the paper is organized as follows. In Section II, we present several background analyses of our research, in Section III, we provide design methodology of our proposed algorithm, pseudo code for proposed algorithm, in Section IV, we provide the result and comparative analysis, in Section V, has the concluding remarks.

\section{Background}

Group Formation is a complex and important step to design effective collaborative learning activities. \parencite[]{10.1007/978-3-319-10166-8_18}. In 2014, Wilmax Marreiro Cruz and Seiji Isotani were used Computer-Supported Collaborative Learning (CSCL) and they foucs on developing and testing group formation in collaborative learning contexts using best practices and other pedagogical approaches.They use CSCL context to fill up the gap of the complexity of that. Initially, they searching on six digital libraries, they collected 256 studies. Then, after a careful analysis of each study, they verified that only 48 were related to group formation applied to collaborative learning contexts. Finally, they categorized the contributions of their study to present an overview of the findings produced by the community.

In 2019, Anna Sapienza at al. \parencite[]{10.3389/fdata.2019.00014} were trying to predict team using Deep Neural Network. They mostly emphasis the social impact and online game. They collect their data from Dota2 game and they generate a directed co-play network, whose links' weights depict the effect of teammates on players' performance. Specifically, they propose a measure of network influence that captures skill transfer from player to player over time. \parencite[]{10.3389/fdata.2019.00014}. They finally provide insights into skill transfer effects: their experimental results demonstrate that such dynamics can be predicted using deep neural networks.

In 2019, Soheila Garshasbi at al. \parencite[]{GARSHASBI2019506} they applied their algorithm's to find the optimal group on education system. In their paper, they propose a novel algorithm capable of properly addressing a variety of optimization problems in optimal learning group formation processes. To this end, a multi-objective version of Genetic Algorithms, i.e. Non-dominated Sorting Genetic Algorithm, NSGA-II, was successfully implemented and applied to improve the performance and accuracy of optimally formed learning groups. There approach is not only application for the education system but also work in other domain as well.

In 2019, Kaj Holmberg \parencite[]{doi:10.1080/01605682.2018.1500429} created a software where within 60 seconds a group formed. In his research, he used heuristics and metaheuristics for the problem. Computational tests are made on randomly generated instances as well as real life instances. Some of the heuristics give good solutions in short time. 

In 2010, Kalliopi Tourtoglou and Maria Virvou \parencite[]{Tourtoglou2010}, in their book chapter, they describe between local search and Simulated Annealing (SA) and also describe the results of those. The local optimization algorithms start with an initial solution and repeatedly search for a better solution in the neighborhood with a lower cost. So, the locally optimal solution to which they result is this with the lower cost in the neighborhood. On the other hand, SA is motivated by the desire to avoid getting to trap on local optima. They also emphasize Computer-Support Collaborative Learning (CSCL) process that is improve teaching and learning with the help of modern information and communication technology.

\section{Methodology}
\subsection{Data Selection}
The dataset we are using contains 818 students, where
the number of students is enrolled in the Southeast University, Dhaka, Bangladesh. Originally the data set has Student ID, Course Code, Total, Credits, Semester, Gender, Marks Round,	Grades. We take one consecutive course result and their prerequisite course and try to predict the best group of maximum 3 members of each group. Around 69 distinct courses and 818 students so altogether 36833 rows in our dataset. We also have prerequisite course data.

\subsection{Data Preparation}
First, we sort the data according to Student ID, Course Code, Semester and Marks Round.Then we drop duplicates for Student ID and Course Code and Keep only the last value because we need only the highest grade of a student if he did the course more than one time for a good grade. After that we drop Total and Grade Column because we have alternative Marks Round column. Then we also drop those rows that contain marks less than 40 and by doing that we ensure the dataset has not any rows that contain failed information of a course. At the end the dataset contains 23382 rows and attributes 	Student ID, Course Code, Credits, Semester, Gender, Marks Round . We keep only the Course code and Prerequisite Column from the Prerequisite dataset.

\begin{center}
\begin{table}[h!]
\begin{tabular}{|p{2cm} | p{5cm}|}
\hline
Attributes & Remarks\\ 
\hline
Student ID & Identiti of a student\\ 
\hline
Course Code & Students are grouped for a specific course\\
\hline
Credit & For define it is theory or lab course\\ 
\hline
Semester & Semester also important for clustering\\ 
\hline
Marks Round & Total marks for a course\\ 
\hline
Prerequisite & To find out previous record in a particular student\\ 
\hline
\end{tabular}
\end{table}
\end{center}
We separated 30 students (in a class or section) and grouped them into 10 different group with arbitrary cluster, based on above table.

\subsection{Algorithm} 
\subsubsection{Simulated Annealing}
This module provides a hyperparameter optimization using simulated annealing. It has a SciKit-Learn-style API and uses multiprocessing for the fitting and scoring of the cross validation folds. The benefit of using Simulated Annealing over an exhaustive grid search is that Simulated Annealing is a heuristic search algorithm that is immune to getting stuck in local minima or maxima.
\subsubsection{Details of Simulated Annealing Algorithm}
\begin{itemize}
    \item Start with some Initial T and alpha
    \item Generate and score a random solution (score\_old)
    \item Compare score\_old and score\_new :
    \begin{itemize}
        \item if score\_new $>$ score\_old: move to neighbour solution
        \item if score\_new $<$ score\_old: maybe move to neighbour solution
    \end{itemize}
    \item Decreases T: T*=alpha
    \item Repeat the above steps until one of the stopping coditions met: 
    \begin{itemize}
        \item T $>$ T\_min
        \item n\_iterations $>$ max\_iterations
        \item total\_runtime $>$ max\_runtime
    \end{itemize}
    \item Return the score and hyperparameters of the best solution
\end{itemize}
The decision to move to a new solution from an old solution is probabilistic and temperature dependent. Specifically, the comparison between the solutions is performed by computing the acceptance probability
\begin{equation}
    A = \exp(\frac{score\_new - score\_old}{T})
\end{equation}
The value of A is then compared to a randomly generated number in [0,1]. If A is greater than the randomly generated number, the algorithm moves to the hyperparameters of the neighboring solution. This means that while T is large, almost all new solutions are preferred regardless of their score. As T decreases, the likelihood of moving to hyperparameters resulting in a poor solution decreases.
\subsubsection{Importance of Regarding Scoring}
This implementation of Simulated Annealing can use any of the built-in SciKit Learn scoring metrics or any other scoring function/object with the signature score(estimator, X, Y). It is important to note that during the annealing process, the algorithm will always be maximizing the score. So if you intend on finding optimal hyperparameters for a regression algorithm, it is important to multiply your scoring metric by -1.
\subsubsection{Select a Cooling Schedule}
While there are lots of researchers looking into best practices for selecting a cooling schedule, I've had good results with the following practices. Early on, it's good if the algorithm is pretty indiscriminate. To achieve this, the acceptance probability should be close to or greater than 1 for all values of $score\_new - score\_old$. For scoring metrics that take on values [0, 1], this means setting $T >> 1$. However, you don't want the algorithm to spend too much time with the temperature this high because it doesn't care much about moving towards better solutions. Thus, for initial temperatures that are high, use values of alpha that will result in rapid cooling: $0.5 < alpha < 0.8$. Lastly, with a rapid cooling schedule, select a $T\_min$ that is low enough to properly explore the input hyperparameter space, e.g $T\_min = 0.0001$. To calculate the number of steps in the cooling schedule use:
\begin{equation}
    K = \frac{\log(T\_min)-\log(T)}{\log(alpha)}
\end{equation}

\subsection{Pseudo Code}
Implementation of the algorithm was done with the help
of following tools: Python, Pandas, Numpy, Scikit-learn, matplotlib, Google Colab. 

\begin{algorithm}

\SetKwData{best_score}{best_score}
\SetKwData{best_data}{best_data}
\SetKwData{X_train}{X_train}
\SetKwData{X_test}{X_test}
\SetKwData{y_train}{y_train}
\SetKwData{y_test}{y_test}
\SetKwData{clf}{clf}

\SetKwFunction{DataFrame}{DataFrame}
\SetKwFunction{train_test_split}{train_test_split}
\SetKwFunction{LinearSVC}{LinearSVC}
\SetKwFunction{temperature}{temperature}
\SetKwFunction{neighbour}{neighbour}

\SetKwInOut{Input}{input}

\Input{Student`s record and prerequisite course record}

\BlankLine
\emph{Separate 30 students with the specific semester and course}\;
\emph{best\_score = 0.0}\;
\emph{best\_data = pd.DataFrame()}\;
\For{$i\leftarrow 0$ \KwTo $n$}{
    \emph{Cluster 30 students with maximum 3 members}\;
    \emph{Then split the data with $sklearn.model_selection$}\;
    \emph{$X\_train, X\_test, y\_train, y\_test = train\_test\_split(X,y,test\_size=0.2)$}\;
    \emph{Initialize linear SVM classifier}\;
    \emph{$clf = svm.LinearSVC()$}\;
    \emph{Pass $clf$ as parameter of Simulated Annealing}\;
    
}
\caption{Pseudo Code}\label{algo_disjdecomp}
\end{algorithm}

\section{Result \& Analysis}
\subsection{Different result analysis}
We use several iterations to get the best result with success and error rate. For example, we fixed the iteration into 5 times. Each iteration we get a different success rate and error rate.
\begin{itemize}
    \item Iteration 1: On the 1st iteration we got 50\% to 70\% success rate, that is not good enough. But we save the result as a current and best score.
    
    \item Iteration 2: On the 2nd iteration it gives us 75\% to 87\% success rate, now we can say it is doing better than before. Then we save the result in a current score and  compare it with iteration 1. Iteration 2 gives more accuracy than before so that we save it to the best score.
    
    \item Iteration 3: On the 3rd iteration it goes the highest of all iteration that is 97\% to 99\%. According to our algorithm process we save it to the current score and compare it with the previous best score that we have. We found it higher than previous so that we save it to the best score.
    
    \item Iteration 4: On the 4th iteration, our model gives us 92\% to 95\% success rate. Following the same process we save it to the current score but we also don’t save it to the best score because it is not higher from the previous best score.
    
    \item Iteration 5: On the last and 5th iteration, we can see that our proposed model gives us a success rate that belongs to 94\% to 97\%. In this iteration we get better result than iteration 4 but still now it can’t overtake the accuracy result of iteration 3.
    
\end{itemize}
Finally, our accuracy upon 30 students, now has been listed in the table. In this table we can clearly see what actually happens while iteration goes on.

\begin{center}
\begin{tabular}[h!]{ |c|c| } 
 \hline
 Iterations  & Accuracy Score\\
  \hline
 Iteration 1 & 66.78\%\\ 
 Iteration 2 & 86.34\%\\ 
 Iteration 3 & 98.12\%\\
 Iteration 4 & 93.42\%\\
 Iteration 5 & 97.17\%\\
 \hline
\end{tabular}
\end{center}

\subsection{Comparison Analysis}
In the part we will discuss our proposed model result with the actual one.\\

    \begin{figure}[h!]
        \centering
        \includegraphics[width=\linewidth]{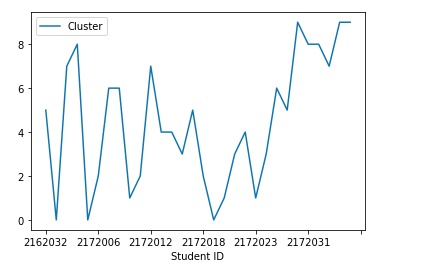}
        \caption{Cluster Plot Before Optimization}
        \label{fig: Cluster Plot Before Optimization}
    \end{figure}
    
In Fig-1 , it shows us the arbitrary cluster plot visualization where we can see that the line is in an non-linear position.\\
    \begin{figure}[h!]
        \centering
        \includegraphics[width=\linewidth]{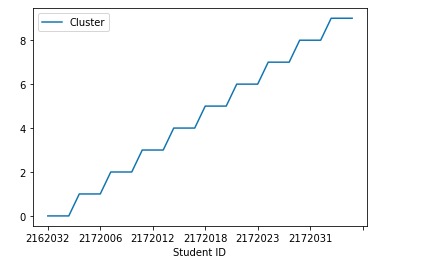}
        \caption{After Optimization}
        \label{fig: After Optimization}
    \end{figure}
In Fig-2 after all iteration and optimizing with our proposed algorithm and taking the best accuracy we get the optimal group formation in a quite linear position. Where the success rate is 97 to 99\%. In our case it give us 98.12\% among 30 students.
 .

\section{Conclusion}
Our primary objective is to find the optimal group based on students previous academic history. In our analysis it confirms that previous performance have a significant impact over students` current performance. Again for irregular students, it`s hard to predict their actual weight. 
Our research goal was to propose a generate groups using SA which will be suitable for predicting the students group in class. Algorithm similar to the one developed, as well as any enhancements thereof may become an incorporated area for many academic institution.

\printbibliography

\end{document}